\documentclass[conference]{IEEEtran}
\IEEEoverridecommandlockouts

\usepackage{graphicx}     
\usepackage{amsmath}      
\usepackage{eurosym}
\usepackage{balance}
\usepackage{paralist}
\usepackage{blindtext}
\usepackage{xcolor,colortbl}
\usepackage{soul}
\usepackage{comment}
\usepackage{todonotes}

\usepackage{mdframed}
\usepackage{enumitem}
\usepackage{multirow}

\newenvironment{itemize*}%
  {\begin{itemize}%
    \setlength{\itemsep}{0pt}%
    \setlength{\parskip}{0pt}}%
  {\end{itemize}}
  
\usepackage{array}
\usepackage{longtable}

\definecolor{Gray}{gray}{0.95}
\definecolor{LightGray}{gray}{0.55}
\definecolor{LightCyan}{rgb}{0.88,1,1}
\definecolor{battleshipgrey}{rgb}{0.52, 0.52, 0.51}
\definecolor{Cyan}{rgb}{0.0, 0.72, 0.92}

\newcolumntype{L}[1]{>{\raggedright\let\newline\\\arraybackslash\hspace{0pt}}m{#1}}
\newcolumntype{C}[1]{>{\centering\let\newline\\\arraybackslash\hspace{0pt}}m{#1}}
\newcolumntype{R}[1]{>{\raggedleft\let\newline\\\arraybackslash\hspace{0pt}}m{#1}} 
\newcolumntype{g}{>{\columncolor{Gray}}p}

\usepackage{enumitem}

\usepackage[bookmarksopen,
     colorlinks,
     linkcolor=black,
     urlcolor=black,
     citecolor=black,
     breaklinks,
     linktocpage=true]{hyperref}
\AtBeginDocument{%

}

\usepackage[nomain,acronym]{glossaries}
\makeglossaries

\newcommand{\singular}[1]{\gls{#1}\xspace}

\newcommand{\abkuerzung}[2]{\newacronym{#1}{#1}{#2}}

\abkuerzung{AADL}{Architecture Analysis \& Design Language}

\abkuerzung{ADL}{architecture description language}

\abkuerzung{ADR}{Action Design Research}

\abkuerzung{API}{application programming interface}

\abkuerzung{AS}{abstract syntax}

\abkuerzung{AAS}{Asset Administration Shell}

\abkuerzung{AST}{abstract syntax tree}

\abkuerzung{CBSE}{component-based software engineering}

\abkuerzung{CD}{Class Diagram}

\abkuerzung{CFG}{context-free grammar}

\abkuerzung{CoCo}{context condition}

\abkuerzung{CS}{concrete syntax}

\abkuerzung{CPS}{cyber-physical system}

\abkuerzung{DoE}{design of experiments}

\abkuerzung{DSL}{domain-specific language}

\abkuerzung{DTC}{Digital Twin Consortium}

\abkuerzung{FSM}{finite-state machine}

\abkuerzung{GPL}{general-purpose programming language}

\abkuerzung{IDTA}{Industrial Digital Twin Association}
\newcommand{\idta}{\singular{IDTA}}

\abkuerzung{KR}{knowledge representation}

\abkuerzung{LCDP}{low-code development platform}

\abkuerzung{DPL}{DSL product line}

\abkuerzung{M2M}{model-to-model}

\abkuerzung{M2T}{model-to-text}

\abkuerzung{MBSE}{model-based systems engineering}

\abkuerzung{MDE}{model-driven engineering}

\abkuerzung{MDSE}{model-driven systems engineering}

\abkuerzung{OCL}{Object Constraint Language}

\abkuerzung{SC}{statechart}

\abkuerzung{Sem}{semantics}

\abkuerzung{SME}{small and medium-sized enterprise}

\abkuerzung{SLE}{software language engineering}

\abkuerzung{SOS}{Structural Operational Semantics}

\abkuerzung{SPL}{software product line}

\abkuerzung{SysML}{Systems Modeling Language}

\abkuerzung{TS}{technological space}

\abkuerzung{TSCA}{time-synchronous channel automata}

\abkuerzung{TSPA}{time-synchronous port automata}

\abkuerzung{UML}{Unified Modeling Language}

\usepackage{xspace}
\usepackage{ifdraft}

\newcommand{\pt}{AS\xspace}
\newcommand{\pts}{ASs\xspace}

\newcommand*{\ie}{\textit{i.e.,}\@\xspace}
\newcommand*{\eg}{\textit{e.g.,}\@\xspace}
\newcommand*{\cf}{\textit{cf.}\@\xspace}

\makeatletter
\newcommand*{\etc}{%
  \@ifnextchar{.}%
  {\textit{etc}}%
  {\textit{etc.}\@\xspace}%
}
\makeatother

\definecolor{se-green}{RGB}{0,128,0}
\definecolor{se-blue} {RGB}{0,0,204}

\newcommand{\code}[1]{\texttt{#1}}
\newcommand{\cw}[1]{\code{#1}}

\newcounter{papernumber}

\newcounter{requirement}[section]

\definecolor{MyBoxText}{RGB}{255,255,255}
\definecolor{MyBoxBG}{RGB}{13,50,153}
\definecolor{MyBoxText}{RGB}{0,0,0}
\definecolor{MyBoxBG}{RGB}{255,255,255}

\newcounter{conclusion}[section]

\newcommand{\conclusionBox}[2]{
	\stepcounter{requirement}
	\begin{mdframed}[linewidth=0.3mm, backgroundcolor=MyBoxBG, rightline=true, leftline=true, innertopmargin=2mm, innerbottommargin=2mm, innerleftmargin=2mm, innerrightmargin=2mm]
	\color{MyBoxText}
	{\sffamily
	\textbf{Conclusion C#1:} #2 
	}
	\end{mdframed}
}

\makeatletter
\newcommand{\linebreakand}{%
  \end{@IEEEauthorhalign}
  \hfill\mbox{}\par
  \mbox{}\hfill\begin{@IEEEauthorhalign}
}
\makeatother

\usepackage{ifdraft}
\usepackage{ifthen}

\newboolean{debug} 
\setboolean{debug}{false}
 
\ifthenelse{\boolean{debug}}{
  \usepackage{showframe} 
  \usepackage{todonotes} 
  \newcommand{\xynote}[2]{\todo[inline]{#1: #2}}
  
  \newcommand{\smallnote}[2]{
    \par\noindent\makebox[\textwidth][c]{%
      \fbox{
        \begin{minipage}{0.9\textwidth}
          \scriptsize{\color{#1}{#2}}
        \end{minipage}
      }
    }
  }
  \usepackage{lineno}
  \linenumbers
  
  \newcommand{\copyandpaste}[1]{{\color{red}#1}}
  \newcommand{\hint}[1]{\smallnote{black}{#1}}
  \newcommand{\thought}[1]{\smallnote{teal}{#1}}
  \newcommand{\actionitem}[2]{\noindent$\circ$ #1 (#2)\\}
  \newcommand{\xactionitem}[2]{\noindent$\times$ #1 (#2)\\}
  
}{%
  \hypersetup{hidelinks} 
  \newcommand{\xynote}[2]{}
  \newcommand{\smallnote}[1]{}
  
  \newcommand{\hint}[1]{}
  \newcommand{\thought}[1]{}
  \newcommand{\actionitem}[2]{}
  \newcommand{\xactionitem}[2]{}
  \newcommand{\copyandpaste}[1]{}
  
}

\DeclareUnicodeCharacter{301}{XXXXXXXXXXX}
\DeclareUnicodeCharacter{302}{XXXXXXXXXXX}
\DeclareUnicodeCharacter{308}{XXXXXXXXXXX}

\usepackage{listings}
\newenvironment{ownfigure}[0]%
{\begin{figure}[htb!]}
{\end{figure}}

\usepackage{color}
\definecolor{DarkRed}{rgb}{0.75,0,0}
\definecolor{Lightgreen}{rgb}{0.588,1.0,0.588}
\definecolor{DarkGreen}{rgb}{0,0.5,0}
    
\lstdefinelanguage{MontiArc}[]{Java}{
  morekeywords={component, port, in, out, inv, package, import, connect, autoconnect}
}

\lstdefinelanguage{myJava}[]{Java}{
  commentstyle=\color{DarkGreen}\itshape 
}

\lstdefinelanguage{MontiArcAutomaton}[]{Java}{
  morekeywords={component, port, in, out, inv, package, import, connect,
  autoconnect, automaton, state, ocl, java, initial, final,
  noCompletion, chaosCompletion, var, mode, activate, transitions,
  modetransitions}, commentstyle=\color{DarkGreen}\itshape }

\lstdefinelanguage{MCConfig} { 
    morekeywords={config, Require, Model} 
}

\lstdefinelanguage{Manifest} { 
    morekeywords={Manifest, Bundle, ManifestVersion, Name, SymbolicName,
      Version, Require
    } 
}

\lstdefinelanguage{mcGrammar}[]{}{
  morekeywords={
    grammar, package, path, parser, lexer, nows, noslcomments, nomlcomments, 
    noident, nostring, noanything, nocharvocabulary, dotident, identrule,
    xmlcomments, hashcomments, texcomments, freemarkercomments, concept, 
    globalnaming, define, usage, options, true, false, protected, ident, 
    compilationunit
  }
}

\lstdefinelanguage{mcLng}[]{}{
  morekeywords={
    dsltool, language, package, path, parser, root, parsingworkflow, 
    rootfactory, lexer, nows, noslcomments, nomlcomments, noident, nostring,
    dotident, concept, globalnaming, define, usage, options, true, false, 
    protected, ident
  }
}

\lstdefinelanguage{mcManifest}[]{}{
  morekeywords={
    bundle, Bundle, Name, SymbolicName, true, false, Main, Class, 
    Version, Activator, Localization, Require, 
    Exclude, Eclipse, LazyStart, Vendor, Export, Package, 
    ClassPath
  }
}

\lstdefinelanguage{Alloy}[]{Java}{
commentstyle=\color{DarkGreen}\itshape,
  morekeywords={abstract,sig,->,fact,pred,fun,run,for,iff,
  not,no,one,all,some,lone,\#,set,in,and,or,but,exactly,none,univ,Int,assert,check},
  otherkeywords = {[2]????},
    morekeywords = {[2]????},
    keywordstyle={[2]\color{blue}},
    otherkeywords = {[3]????,<,<->,->, &, |, =, !=, !,<:,~},
    morekeywords = {[3]????,<,<->,->, &, |, =, !=, !,<:,~},
    keywordstyle={[3]\color{blue}}
  }

\lstdefinelanguage{mccd}[]{Java}{
  morekeywords={classdiagram,abstract,<<singleton>>,class,int,String,
  association,composition,extends}
}

\lstdefinelanguage{FreeMarker}[]{}{
  keywordsprefix={\#},
  keywords={in},
  commentstyle=\color{DarkGreen}\itshape }

\lstdefinelanguage{Mona}[]{}{
  morekeywords={ex0,all0,ex1,all1,ex2,all2,var0,var1,var2,pred,in,notin,include,union,inter,empty,assert},
  morecomment=[l]{\#},
  commentstyle=\color{DarkGreen}\itshape,
  otherkeywords = {[2]????,next,boolean,init,case,esac},
  morekeywords = {[2]????,next,boolean,init,case,esac},
  otherkeywords = {[3]????,<,<=>,=>, &, |, =, !=, !},
  morekeywords = {[3]????,<,<=>,=>, &, |, =, !=, !},
}

\lstdefinelanguage{myPython}[]{Python}{
  morekeywords={assert},
  morecomment=[l]{\#},
  commentstyle=\color{DarkGreen}\itshape,
}

\lstdefinelanguage{GeneratorConfiguration}[]{Java} {
  morekeywords={
    template, 
    generator, 
    ast, 
    runtime},
}

\lstdefinelanguage{ApplicationConfiguration}[]{Java} {
  morekeywords={
    application,
    behaviors,
    bindings,
    classdiagrams,
    components,
    factories,
    generators,
    map,
    to},
}

\lstdefinelanguage{Isabelle}[]{} {
    morekeywords={
        datatype,
        typedef},
}

\usepackage{cite}
\usepackage{amsmath,amssymb}
\usepackage{algorithmic}
\usepackage{graphicx}
\usepackage{textcomp}
\usepackage{rotating}
\usepackage{tikz}
\usepackage{anyfontsize}

\def\BibTeX{{\rm B\kern-.05em{\sc i\kern-.025em b}\kern-.08em
    T\kern-.1667em\lower.7ex\hbox{E}\kern-.125emX}}

\definecolor{tableHeaderColor}{rgb}{0.63, 0.79, 0.95}
\definecolor{tableHeaderColor}{rgb}{0.0, 0.5, 1.0}

\begin{document}

\title{Towards a Unifying Reference Model for Digital Twins of Cyber-Physical Systems
\thanks{Funded by the Deutsche Forschungsgemeinschaft (DFG, German Research Foundation) -- Model-Based DevOps -- 505496753. -- Funded by the Agence Nationale De La Recherche (ANR) – France --
Website: \url{https://mbdo.github.io}. }}

\author{\IEEEauthorblockN{J\'{e}r\^{o}me Pfeiffer}
\IEEEauthorblockA{\textit{University of Stuttgart} \\
Stuttgart, Germany \\
jerome.pfeiffer@isw.uni-stuttgart.de}
\and
\IEEEauthorblockN{Jingxi Zhang}
\IEEEauthorblockA{\textit{University of Stuttgart} \\
Stuttgart, Germany \\
jingxi.zhang@isw.uni-stuttgart.de}
\and
\IEEEauthorblockN{Benoit Combemale}
\IEEEauthorblockA{\textit{University of Rennes} \\
Rennes, France \\
benoit.combemale@irisa.fr}

\linebreakand

\IEEEauthorblockN{Judith Michael}
\IEEEauthorblockA{\textit{RWTH Aachen University} \\
Aachen, Germany \\
michael@se-rwth.de}
\and
\IEEEauthorblockN{Bernhard Rumpe}
\IEEEauthorblockA{\textit{RWTH Aachen University} \\
Aachen, Germany \\
rumpe@se-rwth.de}
\and
\IEEEauthorblockN{Manuel Wimmer}
\IEEEauthorblockA{\textit{JKU Linz} \\
Linz, Austria \\
manuel.wimmer@jku.at}
\and
\IEEEauthorblockN{Andreas Wortmann}
\IEEEauthorblockA{\textit{University of Stuttgart} \\
Stuttgart, Germany \\
wortmann@isw.uni-stuttgart.de}
}









\maketitle              

\begin{abstract}

Digital twins are sophisticated software systems for the representation, monitoring, and control of cyber-physical systems, including automotive, avionics, smart manufacturing, and many more. 
Existing definitions and reference models of digital twins are overly abstract, impeding their comprehensive understanding and implementation guidance. 
Consequently, a significant gap emerges between abstract concepts and their industrial implementations.
We analyze popular reference models for digital twins and combine these into a significantly detailed unifying reference model for digital twins that reduces the concept-implementation gap to facilitate their engineering in industrial practice.
This enhances the understanding of the concepts of digital twins and their relationships and guides developers to implement digital twins effectively. 


\end{abstract}

\section{Introduction}

Research and industry employ digital twins (DTs)~\cite{kritzinger2018digital,tao2018digital} for various kinds of cyber-physical systems (CPSs), in 
many domains, including automotive, construction, energy management, medicine, smart manufacturing, and more~\cite{DJR+22,MBD+24}.
They recently have been standardized in manufacturing~\cite{tc1842020iso} to become a central part of automated manufacturing systems.
Yet, despite various definitions, there is little consensus about what a DT actually is and what it should comprise.
To mitigate this, research and practice have produced various reference models of DTs, which aim to explain their main constituents and relations.
Advancing the engineering of DTs requires establishing a detailed and comprehensive grasp of their constituents and their relationships.
To guide their systematic engineering, we present a unifying reference model of DTs of greater detail that helps reduce the concept-implementation gap by introducing consistent descriptions of common abstract concepts proposed in the existing reference models.  
Our contributions to the engineering of DTs, hence, are
\begin{inparaenum}[(1)]
  \item an analysis of reference models of DTs from different domains, that we selected based one popularity, and 
  \item a synthesis of a detailed reference model for DTs that reduces the concept-implementation gap. 
\end{inparaenum}

\section{Reference Models of Digital Twins}
\label{sec:RelatedModels}

\subsection{Reference Models Focusing on Data Flows}
A very popular characterization of DTs~\cite{kritzinger2018digital}, which subsumes the definition by Grieves~\cite{grieves2005product}, distinguishes them from digital models and digital shadows according to the data flows between the actual system (AS), \ie the CPS, and the digital object~\cite{kritzinger2018digital}: 
\begin{inparaenum}[(1)]
\item \emph{Digital model}: If there are no automated data flows between the AS and the digital object, the digital object is considered a digital model, such as a CAD model of a car.
\item \emph{Digital shadow}: If there is an automated data flow from the AS to the digital object, the digital object is called a digital shadow (in the sense that the digital object automatically follows certain changes in the physical object), such as the representation of various properties of interest, \eg mileage, speed, on a modern car's digital dashboard.
\item \emph{Digital twin}: If there is an automated data flow back from the digital object to the AS as well, the digital object is considered a DT, as changes on the digital object then also entail changes on the AS.
\end{inparaenum}
And while these categories enable a qualitative distinction between digital models, digital shadows, and DTs, they are at a high level of abstraction as they omit to specify any details on the constituents of DTs. 

\begin{table*}[t]
\caption{Comparison of the main concepts within the discussed reference models of DTs. 
}
\centering
\setlength\tabcolsep{1.5pt}
\renewcommand{\arraystretch}{1.5} 
\begin{tabular}{|g{3.2cm}|p{2.6cm}|p{3.9cm}|p{3.9cm}|p{3.9cm}|}
\hline
\rowcolor{tableHeaderColor} 
\textcolor{white}{\textbf{Main Concept}} & 
\textcolor{white}{\textbf{Kritzinger et al. \cite{kritzinger2018digital}}} & 
\textcolor{white}{\textbf{Tao et al. \cite{tao2018digital}}} &
\textcolor{white}{\textbf{ISO 23247 \cite{tc1842020iso}}} &
\textcolor{white}{\textbf{IDTA (AAS)~\cite{IDTA}}} 
\\
\hline
\textbf{Physical Object} &
physical object &
physical entity (PE)&
observable manufacturing element &
asset 
\\
\hline
\textbf{Digital Object} &
digital object &
virtual models &
digital representation (DR) &
digital representation from a given viewpoint (AAS) 
\\
\hline
\textbf{Model} &
digital model &
virtual entity (VE, a set of models) &
(subsumed by DR) &
submodel 
\\
\hline
\textbf{Data} &
data flows &
digital twin data &
data collection entity &
data element 
\\
\hline
\textbf{Device Communication} &
data flows &
connection &
device communication &
manifest, AAS interfaces 
\\
\hline
\textbf{Service} &
- &
services (of PE and VE) &
services as part of DT entity &
service 
\\
\hline
\textbf{Digital Twin} &
digital twin &
digital twin &
digital twin &
digital twin 
\\
\hline
\end{tabular}
\label{tbl:table_concepts}
\end{table*}


\subsection{Reference Models Focusing on Components}

The 5D model of DTs~\cite{tao2018digital} proposes that DTs consist of five kinds of elements:
\begin{inparaenum}[(1)]
\item \emph{Physical entities} exist within the physical world and serve as the underlying basis for DTs. 
\item \emph{Virtual models} replicate physical entities, as well as their physical properties and behaviors. 
\item \emph{Digital twin data} stems from the physical entity, services, domain expert knowledge, or a combination of different sources.
\item \emph{Services} provided by DTs offer added-value functions, e.g., simulation, verification, monitoring, optimization, diagnosis, prediction.
\item \emph{Connections} in the DT connect the four components to enable collaboration between them.
\end{inparaenum}


\subsection{Reference Models from Industry}

\subsubsection{ISO 23247} This standard describes a reference model of a DT framework for manufacturing~\cite{tc1842020iso}. 
The functional view of this model comprises three layers of entities that provide functionality over observable manufacturing elements (OMEs).
OMEs are items providing observable properties (\eg a manufacturing plant, a service robot, or staff), and their properties are defined through existing  manufacturing standards. 
The \emph{device communication entity layer} comprises functional elements (FEs) to collect and process data from OMEs, as well as to control and actuate OMEs.
The \emph{DT entity layer} uses the device communication entity to, among other functions, represent, manage, operate, simulate, and maintain the devices observed through the OMEs.
Through the \emph{user interface entity layer}, users and additional services can leverage the DT entity and its potentially built-in service sub-entity to reason about and control the twinned system. 

\subsubsection{Asset Administration Shell}

The \idta is developing the Asset Administration Shell (AAS), a standard and framework to collect all information (models, data, documents) of an asset, \eg a milling machine or a robot, systematically and in a machine-processable fashion. 
The \idta envisions three kinds of AAS~\cite{ZEH+25}: the type 1 AAS consists of documents and models; the type 2 AAS additionally collects operation data from the asset it is connected to (e.g., via OPC UA); and the type 3 AAS has built-in analytics to reason about the asset and control it. 
The type 3 AAS, hence, can be understood as an implementation of the DT definition proposed by Kritzinger et al~\cite{kritzinger2018digital}.

\begin{figure*}[ht]
    \centering
    \includegraphics[width=0.9\textwidth]{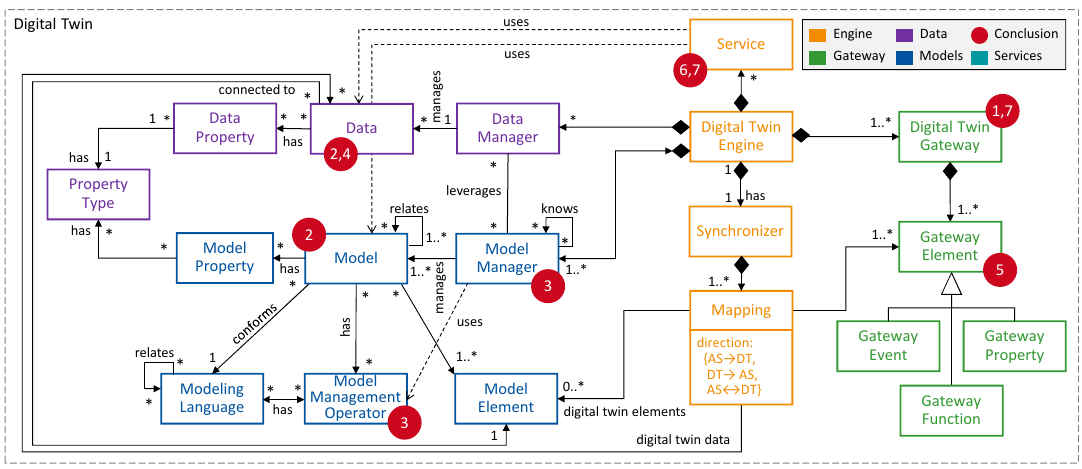}
    \caption{The main DT concepts are a gateway connecting to the \pt, models and data representing it, an engine synchronizing properties of interest between both, and added-value services.}
    \label{fig:04_AW_Conceptual_Model_Full}
\end{figure*}


\subsection{Comparing the Concepts}\label{subsec:related_models_comparison}

We compare the concepts as outlined in~\autoref{tbl:table_concepts}. 
As the reference model of the IDTA~\cite{IDTA} includes concepts from DIN Spec 91345~\cite{DINSpec91345}, we have combined these within one column. 

\subsubsection{Physical Object (according to \texorpdfstring{\cite{kritzinger2018digital,tao2018digital}}))} 

Current approaches for DTs show that the actual system can be anything, from a cyber-physical to a socio-technical, biological, or organizational system. 
Clearly, any physical system itself cannot be part of a \emph{digital} twin. 
Instead, the interfaces and connections between the DT and the AS need to be detailed. 

\conclusionBox{1}{A DT cannot contain its physical twin but needs interfaces to obtain data from and send commands to it.}


\subsubsection{Digital Object (according to~\texorpdfstring{\cite{kritzinger2018digital}}))} 
Opposed to the physical object, the literature refers to its digital representation as, \eg \textit{digital object}~\cite{kritzinger2018digital}, \textit{virtual models}~\cite{tao2018digital}, \textit{digital representation}~\cite{tc1842020iso}, 
or \textit{digital representation from a given viewpoint}~\cite{IDTA}. 
Additionally, the IDTA states that the AAS is the ``standardized digital representation of an asset''~\cite{IDTA}, where an asset is anything of value. 
Harmonizing these terms within a reference model requires further concretization of what such a digital object can be.
Furthermore, the different conceptualizations show that a DT has to include both data as well as models, such that it can be called a DT.

\conclusionBox{2}{Digital objects can be on different abstraction levels ranging from specifically recorded data to general models used for different purposes as is discussed next.}


\subsubsection{Model (according to~\texorpdfstring{\cite{kritzinger2018digital,tao2018digital}}))} 
Kritzinger et al. use the term \textit{digital model}~\cite{kritzinger2018digital} to describe the digital representation of an existing or planned object.
Tao et al. describe the \textit{virtual entity} as a set of \textit{models} (geometry model, physics model, behavior model, and rule model)~\cite{tao2018digital}. 
ISO 23247 uses the term \textit{representation} to describe the ``manner in which information is modeled for interpretation by a machine''~\cite{tc1842020iso}. 
The IDTA uses the term \textit{submodel} as a ``container of SubmodelElements defining a hierarchical structure consisting of SubmodelElements''~\cite{IDTA}. 
Obviously, leveraging models requires the software infrastructure to manage (create, update, delete, search, relate) them among each other and with the data recorded by the DT, \eg to synchronize observed property changes in the twinned \pt with the relevant models the DT has about these properties (and vice versa).    

\conclusionBox{3}{A DT requires various kinds of models. Hence it needs to manage them, and synchronize properties with properties about the twinned system.}


\subsubsection{Data (according to~\texorpdfstring{\cite{kritzinger2018digital,tao2018digital, tc1842020iso,IDTA}}))} 
Kritzinger et al. have a specific focus on \textit{data} and the properties of the data flow, i.e., manual or automatic~\cite{kritzinger2018digital}.
Tao et al. use the term \textit{DT data}, which they denote based on its source (the physical, virtual entity, services, domain knowledge and obtained through data fusion)~\cite{tao2018digital}. 
ISO 23247 uses the term \textit{data}~\cite{tc1842020iso} with a specific focus on its requirements. i.e., data acquisition, analysis, and integrity.
The IDTA defines data elements, where ``a data element is a submodel element that is not further composed of other submodel
elements and has a value. 
This means that DTs will ingest, process, manage, and output various kinds of data (structured data, streams, configurations, \ldots) and need the infrastructure to persist it, perform computations on it, attach metadata, and generally, manage it. 
To make sense of the data, the DT needs to be able to relate data to models.

\conclusionBox{4}{A DT needs to manage different kinds of data and related metadata as well as relate data to its models.}


\subsubsection{Device Communication (according to \texorpdfstring{\cite{tc1842020iso}}))} 
Tao et al. define bi-directional \textit{connections} as a main concept between the different DT constituents~\cite{tao2018digital}.
ISO 23247 refers to \textit{device communication} as a relevant concept between a DT and the observable manufacturing element~\cite{tc1842020iso}, which entails data collection and device control.
The IDTA~\cite{IDTA} refers to the DIN spec 91345 mentioning the term \textit{manifest} as ``externally accessible, defined set of meta information that provides information about the functional and non-functional properties of the I4.0 component''~\cite{DINSpec91345} and DIN EN IEC 63278-1 refers to the \textit{asset administration shell interfaces} as communication provider: 
Hence, the communication between the DT and the \pt is bidirectional, i.e., the DT can send control commands to the \pt.

\conclusionBox{5}{A DT needs to read data from and send a command to its \pt. To be able to react upon changes in the \pt, it also needs to be able to observe changes in the \pt.}


\subsubsection{Service (according to  \texorpdfstring{\cite{tc1842020iso,DTC,IDTA,DINSpec91345}}))}
Tao et al. use the term \textit{services} for the physical entity and the virtual entity~\cite{tc1842020iso}. 
ISO 23247 uses the term \textit{services} as functions provided by the DT entity~\cite{tc1842020iso}. 
The IDTA uses the term \textit{service} with the meaning of a ``demarcated scope of functionality which is offered by an entity or organization via interfaces'' (from~\cite{IDTA} as an extension of~\cite{DINSpec91345}).
Overall, the commonality of the different interpretations of the term service is that everything not vital to the DT's core functionality of representing the \pt digitally can be considered a service (including analytics, AI, etc.).

\conclusionBox{6}{A DT needs means to interface added-value services on top of its core functionality.}

\subsubsection{Digital twin (according to all reference models)} 

All contributions use the term \textit{digital twin} with slightly different definitions, but
most do not explicitly specify where the border of a DT is, \eg \cite{kritzinger2018digital,tao2018digital,IDTA}.

\conclusionBox{7}{A DT needs well-defined boundaries to its environment including the \pt and its services.}

\section{Unifying Digital Twin Reference Model}
\label{sec:ConceptualModel}

Our reference model for DTs (\cf \autoref{fig:04_AW_Conceptual_Model_Full}) is based on the outlined conclusions.


~\\\noindent\textbf{Digital Twin Engine.}\quad
At the heart of the DT is the \cw{Digital Twin Engine} (\cf \autoref{fig:04_AW_Conceptual_Model_Full}), which connects the gateways, the models, the data and the services of the DT.
It comprises a \cw{Synchronizer} over \cw{Mapping}s between the \cw{Model Element}s managed by the DT and the \cw{Gateway Element}s representing the \pts (C3).
A \cw{Mapping} specifies a direction, which can be unidirectional, from \pt to DT or vice versa, or bidirectional (C4).
Moreover, a \cw{Mapping} can be either synchronized at a specified frequency or at the occurrence of a specific trigger.
Overall, the \cw{Digital Twin Engine} connects the \cw{Gateway}s to \cw{Data Manager}s, \cw{Model Manager}s, and eventually \cw{Service}s.


~\\\noindent\textbf{Digital Twin Gateway.}\quad%
To gather data from the twinned \pts directly or from systems observing the twinned \pts, each DT connects to one or more \cw{Gateway}s (C4). 
Each \cw{Gateway} represents a \pt and exposes \pt features of interest as \cw{Gateway Elements} to the DT engine (C1).
These features either are 
\begin{inparaenum}[(1)]
  \item \cw{Properties} that can be read, observed, and synchronized. 
  \item \cw{Events} that the DT engine and the AS may react to. 
  \item \cw{Functions} that can be invoked on the \cw{Gateway} to manipulate the twinned \pt, which can receive arguments and return results, as known from programming.
\end{inparaenum}


~\\\noindent\textbf{Data.}\quad
\cw{Data} has to be collected by the \cw{Digital Twin Engine} through the \cw{Data Manager} that ensures well-formed storage access to specific databases or similar systems (C2). 
Data are associated with some \cw{Data Properties} qualifying it. 
Those properties are of \cw{Property Type}s such as the time (live, historical), if it is raw data or processed one, the origin of the data (e.g., the \pt, or DT services), the uncertainty, the precision, the last update, etc.


~\\\noindent\textbf{Digital Twin Models.}\quad%
If the \cw{Digital Twin Engine} is the heart of the DT, the \cw{Model}s are its brain, which may only be changed through the \cw{Model Manager} or the synchronization of related \cw{Mapping}s. 
Each model conforms to a \cw{Modeling Language} and comprises \cw{Model Properties} of interest, which can belong to different \cw{Property type}s, out of which the latest update is mandatory for DTs supporting bidirectional synchronization (C3).
The languages and the models themselves can have \cw{Model Management Operator}s that govern how the models can be changed while retaining their intra- and inter-model integrity (C3).
The DT engine accesses and manipulates its models only through these \cw{Model Manager}s that are responsible for certain models and embody the language engineering expertise to safely manipulate these.
To this end, each \cw{Model Manager} can delegate changes to other model managers. 
Moreover, the models can be used either offline, \ie not connected to a \cw{Gateway}, \eg to preserve a certain state of the model or to use it for experimentation without affecting the \pts, or online, \ie connected to a \cw{Gateway}, such that changes to the model can affect an \pt and vice-versa (C3)~\cite{tao2018digital}.
Such information is captured as a \cw{Model Property}.


~\\\noindent\textbf{Digital Twin Services.}\quad%
Finally, the \cw{Digital Twin Engine} can purposefully leverage its models and data to provide added-value functionalities through services, such as monitoring key performance indicators, predicting maintenance, or representing the behavior of the twinned system in a specific fashion. 
Therefore, \cw{Service}s can interact with the \cw{Digital Twin Engine} (C5) and, through it, may interact with the \cw{DT Gateway}s. 
However, this interaction may be restricted by the \cw{Digital Twin Engine} (C6) to prevent undesired changes to the DT's models (C3) and the \pt. 

\section{Outlook}
\label{sec:Conclusion}

We analyzed and compared selected reference models from research, industry, and standardization associations. 
Based on these, we devised an initial unifying model that captures the essential concepts of the analyzed models and refines these by detailing some parts of it.
To further detail our reference model, we will investigate (a) further conceptual models of DTs, such as~\cite{DTC,eramo2021conceptualizing,PLW+23}, (b) analyze reference models of services, (c) refine the connection between models and data in our model, and consider the conceptual models implemented by DT platforms~\cite{PfeifferLWW23}.
Then we plan to elaborate on how this unifying model can be translated into DT implementations and underpin this with technological options for each part.
Both will support researchers and practitioners in analyzing and engineering DTs.

\bibliographystyle{ieeetr}
\bibliography{src/bib/main}

\end{document}